\documentclass[twocolumn,aps,superscriptaddress]{revtex4-1}
\usepackage{amsmath,amsfonts,amssymb,amsthm,graphics,graphicx,epsfig,bbm}
\usepackage[colorlinks=true,citecolor=blue,linkcolor=blue,urlcolor=blue]{hyperref}
\usepackage[usenames]{color}
\usepackage{graphicx}
\usepackage{subfigure}
\usepackage{amsmath}
\usepackage{epsfig}
\usepackage{dcolumn}
\usepackage{bm}
\usepackage{color}
\usepackage{times}
\usepackage{epstopdf}
\usepackage{amssymb}
\usepackage{amstext}
\usepackage{latexsym}
\usepackage{hyperref}
\usepackage{amsfonts}
\usepackage{psfrag}
\usepackage{soul,xcolor}
\usepackage[normalem]{ulem}
\usepackage{dsfont}
\usepackage{txfonts}
\usepackage{physics}
\usepackage{footnote}
\usepackage{blindtext}
\usepackage{bbm}
\usepackage{tikz}
\usetikzlibrary{quantikz}
\usepackage{bbold}
\usepackage{siunitx}

\begin{document}
\setstcolor{red}

%\title{Calibrating nanoscale nuclear spin control via NV-diamond correlation spectroscopy}
\title{Coherent Control of Nanoscale Nuclear Spin Ensembles in the Spin Noise Regime
}
%\title{Phase-Controlled Calibration of RF Fields for Reliable NV-Based Nanoscale NMR}
\author{Ana Martin}
\thanks{These two authors contributed equally}
% \email{Corresponding author: ana.martinf@ehu.eus}
\affiliation{Department of Physical Chemistry, University of the Basque Country UPV/EHU, Apartado 644, 48080 Bilbao, Spain}
\affiliation{EHU Quantum Center, University of the Basque Country UPV/EHU, Bilbao, Spain}
\author{Roberto Rizzato}
\thanks{These two authors contributed equally}
\affiliation{Technical University of Munich, TUM School of Natural Sciences, Department of Chemistry, Lichtenbergstra{\ss}e 4, Garching bei M{\"u}nchen, 85748, Germany}
\affiliation{Munich Center for Quantum Science and Technology (MCQST), Schellingstr. 4, M{\"u}nchen, 80799, Germany}
\author{Carlos Munuera-Javaloy}
\affiliation{Institut f\"ur Theoretische Physik und IQST, Universit\"at Ulm, Albert-Einstein-Allee 11, 89081, Ulm, Germany}
\author{Dileep Singh}
\affiliation{Technical University of Munich, TUM School of Natural Sciences, Department of Chemistry, Lichtenbergstra{\ss}e 4, Garching bei M{\"u}nchen, 85748, Germany}
\author{Dominik B. Bucher}
\email{Corresponding author: dominik.bucher@tum.de}
\affiliation{Technical University of Munich, TUM School of Natural Sciences, Department of Chemistry, Lichtenbergstra{\ss}e 4, Garching bei M{\"u}nchen, 85748, Germany}
\affiliation{Munich Center for Quantum Science and Technology (MCQST), Schellingstr. 4, M{\"u}nchen, 80799, Germany}
\author{Jorge Casanova}
\email{Corresponding author: jcasanovamar@gmail.com}
\affiliation{Department of Physical Chemistry, University of the Basque Country UPV/EHU, Apartado 644, 48080 Bilbao, Spain}
\affiliation{EHU Quantum Center, University of the Basque Country UPV/EHU, Bilbao, Spain}

\begin{abstract}

Spin defects in solids, such as the nitrogen-vacancy (NV) center in diamond, have emerged as a key tool for detecting nuclear spins at the nanoscale. While active nuclear spin control via radio-frequency (RF) irradiation is often unnecessary for standard spin-noise detection, it becomes essential for advanced protocols like multidimensional nanoscale NMR. In this work, we investigate nuclear spin control using correlation spectroscopy techniques. We demonstrate, both theoretically and experimentally, that the resulting nuclear spin dynamics depend critically on the initial RF phase and its orientation relative to the NV crystalline axis. Depending on these parameters, identical nuclear rotations can yield full, partial, or even vanishing contrast in the NV readout. These findings highlight a previously underappreciated aspect of spin manipulation in the spin-noise regime: the link between the phase and direction of the applied RF field and its direct impact on correlation-based experiments. Consequently, imperfect calibration of these parameters can lead to ambiguous signal contrasts and misinterpretation of the underlying nuclear spin dynamics. Our results provide deeper insight into nanoscale spin control and pave the way toward reliable multidimensional spin resonance experiments.

%Nanoscale nuclear magnetic resonance employs nitrogen-vacancy (NV) centers as highly sensitive quantum probes for detecting magnetic fields that emanate from nanometer distances. In this work, we investigate synchronized radio-frequency (RF) controls over sample nuclei using correlation spectroscopy techniques. We theoretically show and experimentally demonstrate that accurate detection and characterization of the resulting nuclear spin dynamics, encoded in the NV photoluminescence, depend critically on both the initial phase and the relative orientation between the RF field and the NV crystalline axis. Without proper calibration of the RF parameters --phase and orientation-- signal interpretation become ambiguous, leading to reduced sensitivity and potential mischaracterization of sample properties. Importantly, we show that the effect of the RF field orientation on the NV response can be effectively tuned through controlled variations of the RF initial phase, enabling accurate signal recovery without modifying the experimental geometry. Our results underscore the importance of precise calibration of nuclear spin control parameters to ensure reliable and reproducible quantum sensing at the nanoscale.
\end{abstract}

\maketitle

Spin defects in solids, such as the nitrogen–vacancy (NV) center in diamond, have demonstrated remarkable capabilities for detecting magnetic moments at the nanoscale, down to individual proteins and even single nuclear spins \cite{Devience2015, Lovchinsky2016, Lovchinsky2017, Allert2022, Abendroth2022, Rizzato2023, Du2024}. Unlike conventional bulk nuclear magnetic resonance (NMR), which relies on detecting thermal spin polarization, nanoscale magnetic resonance typically probes stochastic spin fluctuations, often referred to as spin noise \cite{Degen2007, Meriles2010, Muler2013}. This regime offers several advantages: the fluctuating signal scales with the square root of the number of spins, is largely independent of the external magnetic field strength, and does not require active nuclear spin polarization or control.
A central method in nanoscale spin noise detection is correlation spectroscopy \cite{Laraoui2013, Kehayias2017, Degen2017, Liu2021, Liu2022}, in which two dynamical decoupling pulse sequences are applied to correlate temporal fluctuations of the spin noise. To exploit the full capabilities of NMR, particularly the rich toolbox of multidimensional pulse sequences, active nuclear spin control is required \cite{Smits2019, Bruckmaier2023}. Such control would enable, for example, two dimensional nanoscale NMR experiments but only a small number of studies to date have been published in this direction \cite{Boss2016, Aslam2017}. While detecting nuclear spin control is straightforward in the thermal‑polarization regime \cite{Glenn2018, Briegel2025}, calibrating radiofrequency (RF) pulses becomes significantly more challenging in the spin‑noise regime, due to the lack of a coherent NMR signal.

\begin{figure*}[t]
\centering
\includegraphics[width=0.98\textwidth]{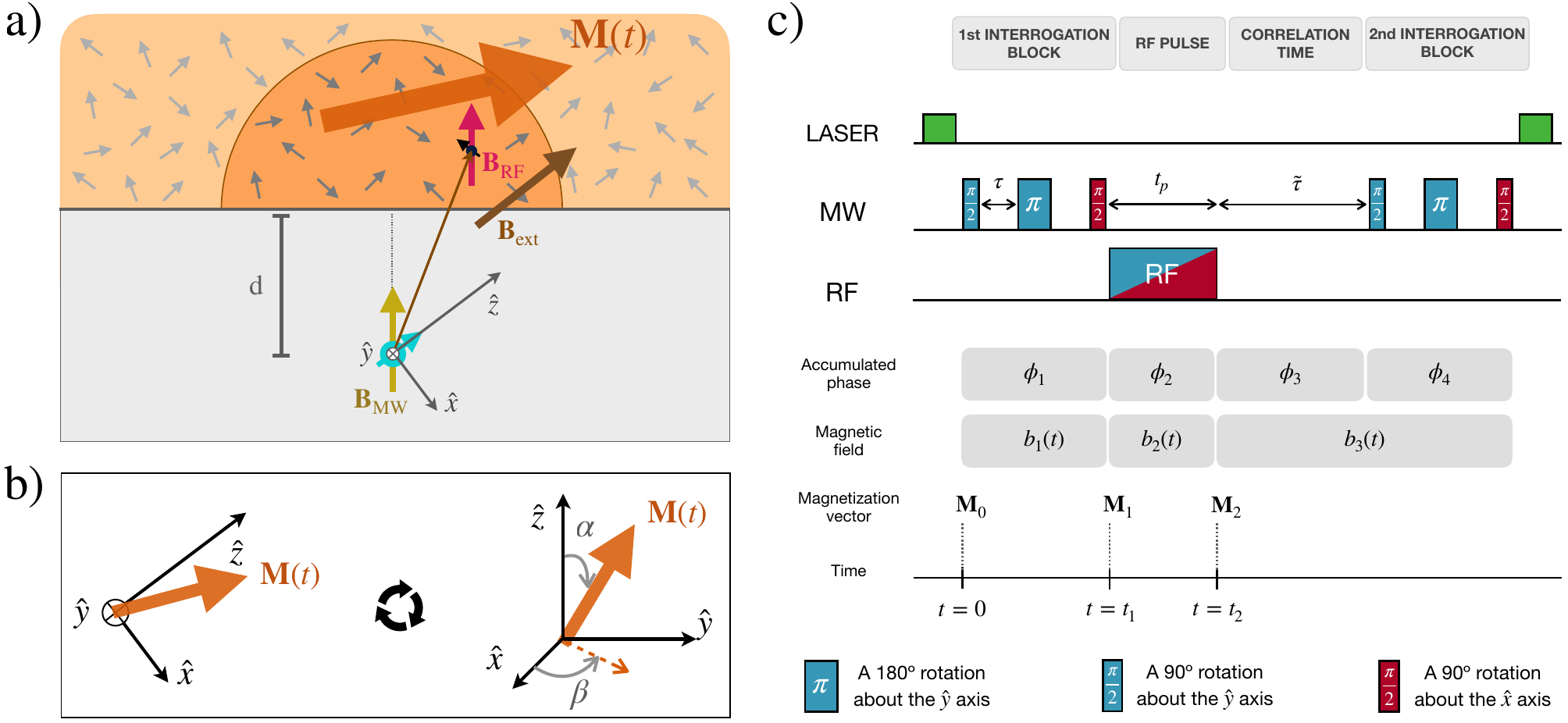}
\caption{{\bf Schematic of the experimental setup and sequence scheme. a)} An organic sample is placed on a diamond hosting an ensemble of shallow NV centers, which act as a nanoscale magnetic sensor. The NV layer is located at a distance $d$ --a few nanometers below the diamond surface-- in close proximity to the sample. ${\bf B}_{\rm ext}$ (brown arrow) is aligned with the NV ($\hat{z}$) axis, while two antennas positioned above the diamond generate the MW (ochre arrow) and RF (magenta arrow) driving fields (${\bf B}_{\rm MW}$ and ${\bf B}_{\rm RF}$ respectively), which are perpendicular to the diamond surface. Each NV center behaves as an independent detector thus, for clarity, the figure shows a single NV and the corresponding region i.e., the detection hemisphere (in dark orange), it senses. {\bf b)} The magnetization vector ${\bf M}(t)$ (orange arrow) of the nuclear ensemble within the NV’s detection hemisphere is shown with respect to the NV axes (left), ${\bf M}(t)$ and the coordinate axes are rotated to better visualize the spherical $\alpha$ and $\beta$ angles (right). {\bf c)} The experimental sequence involves two control channels: a MW channel addressing an NV center and an RF channel applied to the sample. Our protocol consists of four stages: (i) a first interrogation block, (ii) application of an RF pulse on the sample with a phase at $t=0$ that equals $\varphi_{\rm RF}$, (iii) a correlation time interval, and (iv) a second interrogation block. During each stage, the NV accumulates a phase $\phi_i$. The figure also illustrates the magnetic fields $b_i(t)$ detected by the NV at the corresponding steps, the time-evolution of the magnetization vector ${\bf M}(t)$, all together with the timing of the events.}
\label{Fig1}
\end{figure*}

In this work, we investigate theoretically and experimentally the use of RF pulses to coherently manipulate nuclear spin dynamics in the spin noise regime that are read out via correlation spectroscopy, using the experimental setup schematically depicted in Fig.~\ref{Fig1}a. Our protocol, illustrated in Fig.~\ref{Fig1}c, consists of two NV interrogation periods during which the NV sensor accumulates phase from the nuclear spin environment, separated by an RF control segment applied to the nuclear spins. Through both simulations and experiments, we show that the resulting correlation signals (which ideally reflect a nuclear Rabi oscillation) strongly depend on the RF phase as well as on the relative orientation between the RF magnetic field and the NV crystalline axis. As a result, imperfect calibration of these RF parameters can lead to ambiguous signal contrasts and misinterpretation of the underlying nuclear Rabi dynamics. These findings underscore the importance of precise RF phase synchronization and geometric calibration to achieve reliable nuclear spin control in correlation-type nanoscale NMR experiments.

%Nanoscale signal detection enables the analysis of single proteins, two-dimensional materials, and even individual protons~\citep{Allert2022}, making it possible to study single cells, neurons, surfaces, and catalytic processes at extremely small volumes~\citep{Schwartz2019}. Among the most promising tools for such applications are quantum sensors based on nitrogen-vacancy (NV) centers in diamond, which offer exceptional magnetic sensitivity under ambient conditions. Despite this potential, achieving high spectral resolution in nanoscale nuclear magnetic resonance (NMR) remains a significant challenge. Recent efforts have focused on overcoming these limitations through sample confinement and advanced signal processing techniques~\citep{Liu2022}. In this work, we address a complementary and critical aspect of nanoscale NMR: the calibration of nuclear spin control using NV centers, which we show to be strongly dependent on the geometric configuration of the system.

We consider a sample placed on top of an NV-ensemble sensor located approximately $5$~nm below the diamond surface and subjected to an external magnetic field ${\bf B} = B_{\mathrm{ext}}\, \hat{z}$ aligned with the NV axis, as illustrated in Fig.~\ref{Fig1}a. The NV centers serve as quantum sensors that detect the magnetic fields generated by the nuclear spins in the sample. For simplicity, each NV in the ensemble is treated as an independent detector governed by the Hamiltonian
\begin{equation}
H = DS_z^2 - \gamma_e B_{\rm{ext}} S_z - \gamma_e\,b_z(t) S_z + \sqrt{2} \,\Omega\, S_x \cos\left(\omega_{\rm{MW}} t -\varphi_{\rm{MW}}\right),
\end{equation}
where $D = (2 \pi)\ \times\ 2.87$ GHz is the zero-field splitting, $\gamma_e = - \ (2 \pi)\ \times\  28.8$ GHz/T is the electron gyromagnetic ratio, ${\bf S} = (S_x, S_y, S_z)$ represents the spin-$1$ operator of the NV center, $b_z(t)$ denotes the magnetic field generated by the sample, while $\Omega_{\rm{MW}}$ and $\omega_{\rm{MW}}$ are the amplitude and frequency of the MW control drivings. Each NV senses nuclear spins within its detection hemisphere, highlighted in Fig.~\ref{Fig1}a, characterized by a nuclear magnetization vector ${\bf M}(t) = (M_x,M_y,M_z)$ that evolves owing to $B_{\rm ext}$ and to the applied RF pulse of amplitude $\Omega$. This evolution is governed by the equations derived in App.~\ref{appA}. The magnetization vector, ${\bf M}(t)$, gives rise to the $b_z(t)$ field experienced by the NV at every stage.

Now, we evaluate how the magnetic field $b_z(t)$ sensed by the NV changes during different stages of the protocol. This results in three distinct fields: $b_1(t)$ during the {\it 1st interrogation block}; $b_2(t)$ during {\it RF pulse application}; and $b_3(t)$ during {\it correlation time} and {\it 2nd interrogation block} stages, see Fig.~\ref{Fig1}c. Note that RF pulse and correlation time stages last $t_p$ and $\tilde\tau$ respectively.
Explicit derivations of $b_1(t)$, $b_2(t)$, and $b_3(t)$ are in App.~\ref{appB}. 

The application of the RF pulse perturbs the sample magnetization vector ${\bf M}(t)$, thereby causing the magnetic fields sensed by the NV in the first and second interrogation blocks to differ. In particular, one can find (see App.~\ref{appB} for a detailed calculation),
\begin{eqnarray}
    b_1(t) &=& B_{\rm{max}} \, \left[R_{\hat{z}}(\omega t) \; {\bf M}_0\right]\cdot \hat{x},\\
    b_2(t) &=& B_{\rm{max}}\,\left[R_{\hat{z}}(\omega t) \,R_{\hat{k}}(\Omega t)\,{\bf M}_1\right]\cdot \ \hat{x},\\
    b_3(t) &=& B_{\rm{max}}\ \left[R_{\hat{z}}(\omega t)\  {\bf M}_2\right]\cdot\ \hat{x},
\end{eqnarray}
where $B_{\rm max}$ is a field amplitude as defined in App.~\ref{appB}, $R_{\hat{z}}(\omega t)$ and $R_{\hat{k}}(\Omega t)$ are rotation matrices around the $\hat{z}$ and $\hat{k} = \cos{\varphi_{\rm{RF}}}\ \hat{x} + \sin{\varphi_{\rm{RF}}}\ \hat{y}$ axes respectively, with $\hat{x}$, $\hat{y}$, and $\hat{z}$ being the NV crystalline axes, and ${\bf M}_{0,1,2}$ refers to the magnetization vector at the start of distinct stages of the protocol (see Fig.~\ref{Fig1}c).  
Note that $\varphi_{RF}$ is the initial phase of the RF field  ${\bf B}_{\rm{RF}} = B_{\rm{RF}}\ \cos{(\omega_{\rm{RF}}t+\varphi_{\rm{RF}})}\ \hat{x}$. We consider only this component because, in our experimental configuration, ${\bf B}_{\rm{RF}}$ is oriented perpendicular to the diamond surface. For completeness, we want to mention that a misalignment in the RF direction would introduce an additional phase in the vector $\hat{k}$ (see Eq.~(\ref{eqA6}) in App.~\ref{appA}), which, as shown below, would change the NV response.

Now, we consider the initial magnetization $ {\bf M}_0$ as
\begin{equation}
    {\bf M}_0=\begin{pmatrix}\sin{\alpha} \cos{\beta}\\
    \sin{\alpha} \sin{\beta}\\
    \cos{\alpha}
    \end{pmatrix},
\end{equation} 
where $\alpha$ $\in$ $[0, \pi]$ and $\beta$ $\in$ [0, 2$\pi$] are the polar and azimuthal angles (see Fig.~\ref{Fig1}b). We note that the final results presented in this section involve averaging over the angles  $\alpha$ and $\beta$ to properly describe the nanoscale regime, in which nuclear spin signals are inherently non–phase-coherent. 

The magnetic fields associated with the 1st and 2nd interrogation blocks are
\begin{eqnarray}
b_1(t) &=& B_{\rm{max}}\; \sin{\alpha} \cos{(\omega t + \beta)},\\
b_3(t) &=& B_{\text{max}}  \left\{ \cos{(\Omega\ t_p/2)}^2\ \sin{(\alpha)} \ \cos{\left[\omega\ (t+t_p)+\beta\right]} \right.\nonumber \\
&& + \sin{(\Omega\ t_p/2)}^2\ \sin{(\alpha)} \ \cos{\left[\beta-2\ \varphi_{\rm{RF}}-\omega(t+t_p)\right]} \nonumber\\
&& \left.+ \sin{(\Omega\ t_p)}\ \cos{(\alpha)}\ \sin{\left[\varphi_{\rm{RF}}+\omega(t+t_p)\right]} \right\}.
\end{eqnarray}
with $t_p$ being the RF pulse time. Notably, the RF initial phase $\varphi_{\rm{RF}}$ is present in $b_3(t)$. Now we focus on the impact of this phase dependence on the measurement outcomes. Specifically, $\phi_1$ and $\phi_4$  are the accumulated phases during the first and second interrogation blocks; $\phi_2$ is acquired during the RF pulse; and $\phi_3$ accumulates in the correlation time stage (see Fig.~\ref{Fig1}c). A detailed derivation of these phases is presented in App.~\ref{appC}. 

Experimentally, these phases manifest in the NV’s emitted signal, encoded through the spin-dependent photoluminescence (PL). The PL is related to the NV expectation value $\langle\sigma_z\rangle$. 
In particular, for an individual NV center, after the sequence in Fig.~\ref{Fig1}c, one finds 
\begin{equation}
\langle \sigma_z \rangle = \cos(\phi_1)\,\cos(\phi_4)\,\sin(\phi_3+\phi_2)\,-\,\sin(\phi_1)\, \sin(\phi_4), \nonumber \\
\end{equation}
which simplifies under the small-angle approximation to
\begin{equation}\label{eq_sz}
\langle\sigma_z \rangle \approx \phi_3\,+\,\phi_2 - \phi_1\, \phi_4.
\end{equation}
To obtain the ensemble-averaged expectation value $\overline{\langle \sigma_z \rangle}$, we integrate (averaged-sum) over all possible orientations of the initial magnetization vector that defines the nuclear semisphere above each NV in the ensemble. This is

\begin{equation}
\overline{\langle\sigma_z\rangle} = \frac{1}{4\pi}\int_0^{\pi}d\alpha\;\int_0^{2\pi} d\beta \; \left(\phi_3+\phi_2-\phi_1 \phi_4\right).
\end{equation}
By substituting the explicit form of the phases $\phi_i$ (see App.~\ref{appC}), the ensemble-averaged NV  response reads
\begin{eqnarray}
\overline{\langle\sigma_z\rangle}  &=& \frac{2\pi\,B_{\rm{max}}^2 \gamma_e^2}{\omega^2}\left\{\sin{\left(\frac{\Omega t_p}{2}\right)}^2\,\cos{\left[2\varphi_{\rm{RF}}+\omega\,(2t_p+\tilde{\tau})\right]}\right.\nonumber\\
&&- \left.\cos{\left(\frac{\Omega t_p}{2}\right)}^2\,\cos{\left[\omega\;(2t_p+\tilde{\tau})\right]}\right\}.
\label{eq10}\end{eqnarray}
It is worth noting that, in the case of RF field misalignment, the previous result remains valid under the substitution $\varphi_{\rm RF} \rightarrow \varphi_{\rm RF} - \arctan\left(\frac{\Omega_y}{\Omega_x}\right)$, where $\Omega_{x,y}$ denote the Rabi frequencies associated with the $x$ and $y$ components of the RF field. Further details are provided in App.~\ref{appA}.

The expression in Eq.~\eqref{eq10} allows us to analyze the NV response for different initial phases of the RF drive.

\begin{figure*}[t]
\centering
\includegraphics[width=0.85\textwidth]{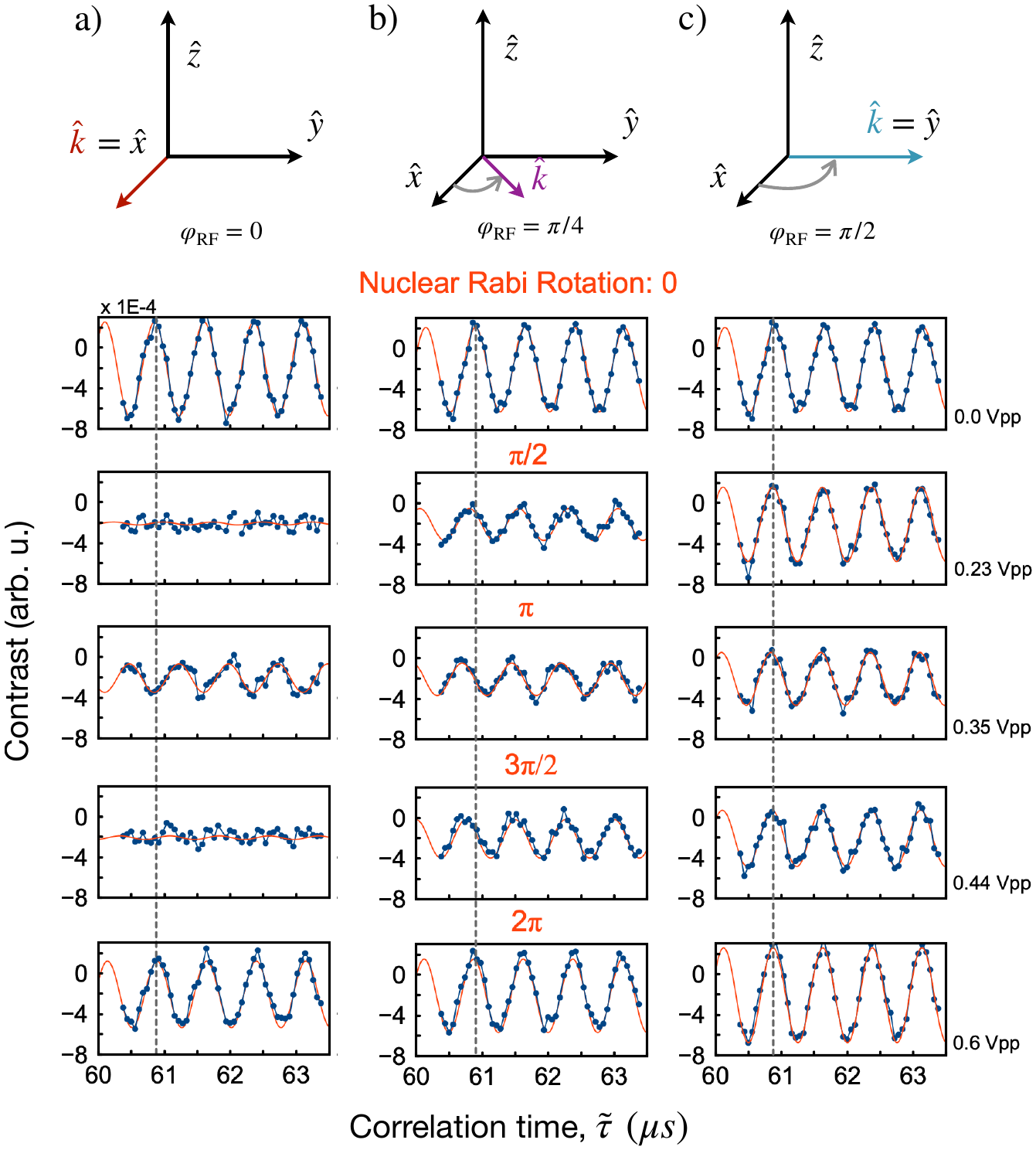} 
\caption{{\bf Effect of RF phase on NV photoluminescence and nuclear spin dynamics.} Theoretical (orange curve, derived from Eq.~\eqref{eq10}) and experimental results (blue dots) showing how the NV PL depends on the initial phase $\varphi_{\rm{RF}}$ of the RF driving. The dashed line serves as a reference to facilitate direct comparison of corresponding evolution times. The effective direction of the RF field (in the rotating frame of the Larmor terms) is defined by the axis $\hat{k} = \cos\varphi_{\mathrm{RF}}\hat{x} + \sin\varphi_{\mathrm{RF}}\hat{y}$ which lies in the plane perpendicular to the NV axis ($\hat{z}$). For each $\varphi_{\rm{RF}}$ ({\bf a}-{\bf c}), the correlation time was swept across several RF amplitudes (top to bottom panels), corresponding to effective spin rotations of $0$, $\pi/2$, $\pi$, $3\pi/2$, and $2\pi$. This allows to identify population inversion under well-calibrated RF driving conditions. {\bf a)} For $\varphi_{\rm{RF}} = 0$, $\hat{k}$ is aligned with $\hat{x}$ and full population inversion occurs. {\bf b)} For $\varphi_{\rm{RF}} = \pi/4$, $\hat{k}$ lies between $\hat{x}$ and $\hat{y}$, producing partial inversion. {\bf c)} For $\varphi_{\rm{RF}} = \pi/2$, $\hat{k}$ is parallel to $\hat{y}$ and no PL contrast is observed. These results demonstrate the strong dependence of the NV response on the RF driving phase and highlight the geometric nature of the phase control.}
\label{Fig2}
\end{figure*}

In particular, for $\varphi_{\rm{RF}}=0$, Eq.~(\ref{eq10}) is
\begin{equation}
\overline{\langle\sigma_z\rangle} =  \frac{2\pi\ B^2_{\rm{max}}\gamma_e^2}{\omega^2}\, \cos{(\Omega t_p)}\, \cos{\left[\omega(2t_p+\tilde{\tau})\right]}.
\end{equation}
In this configuration, the NV photoluminescence depends on the RF Rabi frequency $\Omega$, and clear population inversion is observable at suitable Rabi frequencies, see orange curve in Fig.~\ref{Fig2}a. Conversely, when $\varphi_{\rm{RF}}=\pi/2$, we find
\begin{equation}
\overline{\langle\sigma_z\rangle} = \frac{2\pi\ B^2_{\rm{max}}\gamma_e^2}{\omega^2}\, \cos{(\omega(2t_p+\tilde{\tau}))},
\end{equation}
indicating that the RF pulse has no detectable effect on the NV signal, as it can be observed in Fig.~\ref{Fig2}c). Note that this result holds regardless of the value of the RF amplitude $\Omega$.

For $\varphi_{\rm{RF}} = \pi/4$, $\overline{\langle\sigma_z\rangle}$ takes the form 
\begin{eqnarray}
\overline{\langle\sigma_z\rangle} &=& \frac{2\pi\ B^2_{\rm{max}}\gamma_e^2}{\omega^2}\, \left[\cos{(\Omega t_p)}^2 \cos{(\omega(2t_p+\tilde{\tau}))}\right. \nonumber\\
&&- \left. \sin{(\Omega t_p)}^2 \sin{(\omega(2t_p+\tilde{\tau})}\right],
\end{eqnarray}
leading to partial population inversion, thus to partial PL changes (see Fig.~\ref{Fig2}b). 

These results clearly illustrate that the NV photoluminescence in the spin noise regime depends on the initial RF phase. Furthermore, this dependence implies a direct connection with the NV crystalline axes, as $\varphi_{\rm RF} = 0 \ (\pi/2)$ corresponds to a nuclear rotation around $\hat{x}$ $(\hat{y})$, or around $\hat{k}$ for intermediate values of $\varphi_{\rm RF}$. Consequently, precise control over RF duration and orientation is essential for the reproducibility of NV-based nanoscale quantum sensors.

%--Numerical simulations
Guided by these theoretical insights, we experimentally investigate the PL response of our NV ensemble under different RF driving configurations using a hydrogen-rich sample as a test system. We explore RF-driven nuclear spin dynamics using a correlation-based NV detection protocol, as illustrated in Fig.~\ref{Fig1}c. The measurement consists of two identical XY8-4 dynamical decoupling blocks (four repetitions of the XY8 sequence), which are tuned to the Larmor frequency of $^1\mathrm{H}$ nuclei in a silicon oil layer deposited on the diamond surface. This protocol enables the sensitive detection of statistical nuclear spin fluctuations while suppressing environmental noise.

The two XY8-4 blocks are separated by a variable correlation interval, ${\tilde{\tau}}$, during which nuclear spins evolve freely and are optionally manipulated by an RF pulse. Specifically, a 30\,$\mu$s RF pulse is applied between the two sensing blocks to drive coherent nuclear spin rotations. This RF pulse is preceded and followed by 15\,$\mu$s idle intervals to suppress RF leakage and avoid unwanted interference with the NV control pulses. The correlation time $\tilde\tau$ is subsequently swept to probe the temporal evolution of the nuclear spin ensemble.

To induce nuclear Rabi oscillations, we vary the amplitude of the RF drive while keeping its frequency resonant with the proton Larmor frequency. For each RF amplitude, expressed in volts peak-to-peak (Vpp), a full correlation trace is acquired as a function of $\tilde\tau$ up to 3\,$\mu$s. These traces (Fig.~\ref{Fig2}a) are fitted with a sinusoidal model (see Methods), allowing extraction of the correlation signal amplitude as a function of RF drive strength.

Given a fixed orientation of the RF field relative to the NV crystalline axis --where the $\hat{y}$ component is negligible in our specific geometry (see Fig.~\ref{Fig1}a)-- the initial RF phase becomes the primary experimental parameter to consider.

Microwave control of the NV centers is generated directly by an arbitrary waveform generator (AWG, see Methods), which serves as the master timing source of the experiment and defines the global time origin of each experimental cycle. RF excitation of the nuclear spins is provided by an RF signal generator (see Methods), which is externally triggered by the AWG to produce an RF pulse of defined phase at a certain time within the sequence (after the first XY8-4 block). In other words, if the RF burst is applied at a time \(t=t_1\) (see Fig.~\ref{Fig1}c), the phase of the RF at the moment of burst onset is chosen relative to the chosen value for $\varphi_{\rm RF}$ defined at \(t=0\).

When $\varphi_{\rm RF}=0$ (see Fig.~\ref{Fig2}a), we observe clear signatures of coherent nuclear Rabi oscillations. As the RF amplitude is increased, the correlation signal amplitude first decreases and vanishes, corresponding to an effective \(\pi/2\) nuclear rotation. Upon further increase, the signal reappears with opposite sign, indicating a \(\pi\) rotation of the nuclear spins. With increasing RF amplitude, the signal undergoes a second extinction near \(3\pi/2\) and a subsequent revival near \(2\pi\), consistent with coherent Rabi cycling of the nuclear ensemble.

For  $\varphi_{\rm RF}=\pi/4$ (see Fig.~\ref{Fig2}b) the signal still evolves with RF amplitude, but the inversion is only partial. In contrast, when $\varphi_{\rm RF}$ (see Fig.~\ref{Fig2}c), is set to \(\pi/2\), the correlation signal shows almost no contrast, such that the nuclear spin rotations become effectively invisible in the NV readout despite the presence of RF driving.

In summary, our work identifies the critical role of RF phase and geometric alignment as key variables in correlation-based spectroscopy. By establishing the precise link between these parameters and the resulting signal contrast, we have resolved a source of ambiguity in nanoscale nuclear spin measurements. Our experimental validation confirms that rigorous RF calibration is not merely a technical detail, but a fundamental requirement for the reproducibility of NV-based quantum sensors. This work provides a framework for the robust calibration of coherent nuclear spin control in correlation-type spin noise protocols. This capability is crucial for accurately determining RF drive strengths in (micro)coils and for enabling advanced sensing schemes \cite{Yudilevich2023}, including multidimensional nanoscale NMR experiments \cite{Boss2016}.

%In this work, we have demonstrated theoretically and experimentally the strong dependence of nanoscale quantum sensing signals on the RF phase and geometric alignment between with NV axis. Our findings clearly establish that reliable and reproducible detection of nuclear spin dynamics at the nanoscale requires precise calibration of these parameters. Otherwise signal interpretation is ambiguous, leading to reduced sensitivity and potential misinterpretations of sample properties. Our approach enables accurate signal recovery by simply changing the RF initial phase, paving the way toward improved protocols for nanoscale spectroscopy.

%Compared to previous studies using NV centers for nanoscale magnetic sensing, our results highlight a previously underappreciated aspect -- the link between phase and direction of RF fields in the resulting NV photoluminescence. While many studies focus on pulse optimization and signal processing techniques, our work emphasizes the equally critical role of sensor geometry and RF synchronization. This complements existing literature by addressing a fundamental yet practical challenge in quantum sensing setups.

\begin{acknowledgements}
Authors acknowledge support by the European Union’s Horizon Europe research and innovation programme under Grant Agreement No. 101135742 (QUENCH). J. C. acknowledges the Agencia Estatal de Investigación via the Modelizado, Optimización, y Esquemas de Magnetometria en Centros de Color project PID2024-161371NB-C22, and the Basque Government under Grant No. IT1470-22.
D.B.B acknowledges the Deutsche Forschungsgemeinschaft (DFG, German Research Foundation)—412351169 within the Emmy Noether program and the support by the DFG under Germany’s Excellence Strategy—EXC 2089/1—390776260 and the EXC-2111 390814868. D.S. acknowledges the DFG Walter Benjamin Program, SI 3263/1-1.
\end{acknowledgements}

%--Appendices
\onecolumngrid
\appendix

\section{Dynamics of a spin subjected to RF fields}
\label{appA}
We consider a single nuclear spin in an external magnetic field oriented along the $z$-axis, ${\bf B} = B_{\rm{ext}}\ \hat{z}$ combined with an RF driving field ${\bf B}_{\rm{RF}} = B_{\rm{RF}}\ \cos{(\omega_{\rm{RF}}t+\varphi_{\rm{RF}})}\ \hat{x}$. 
Note that the $\hat{z}$ component of the RF drive does not contribute due to the rotating-wave approximation (RWA), while the RF orientation -perpendicular to the diamond surface- ensures that no $\hat{y}$ component is present.

The Hamiltonian describing the spin dynamics is  given by
\begin{equation}
    H=\frac{\omega}{2}\sigma_z+\Omega\sigma_x\cos{(\omega_{\rm{RF}} t+\varphi_{\rm{RF}})},
\end{equation}
where $\omega = \gamma_n \, B_{\rm{ext}}$ is the nuclear Larmor frequency and $\Omega=\gamma_n B_{\rm{RF}}$ is the Rabi frequency associated with the RF drive, and $\gamma_n$ is the nuclear gyromagnetic ratio.

Going into a rotating frame with respect to $H_0 = \frac{\omega}{2} \sigma_z$ and applying the RWA under the resonance condition $\omega_{\rm{RF}}=\omega$, the Hamiltonian simplifies to
\begin{equation}
    H = \frac{\Omega\sigma_{\varphi_{\rm{RF}}}}{2},
\end{equation}
with
\begin{equation}
    \sigma_{\varphi_{\rm{RF}}} = \begin{pmatrix} 0 & e^{-i\varphi_{\rm{RF}}}\\
    e^{i\varphi_{\rm{RF}}} & 0
    \end{pmatrix}.
\end{equation}
This implies that $H = \frac{\Omega\sigma_{\varphi_{\rm{RF}}}}{2}$ imprints rotations along the axis 
\begin{equation}
\hat{k}= \cos\varphi_{\rm{RF}} \ \hat{x} + \sin\varphi_{\rm{RF}} \hat{y},
\end{equation}
lying on the XY plane.

Before continuing, we analyze the scenario including a miscalibration of the RF field. In particular, we consider the RF having a component $B'_{\rm RF}$ over $\hat{y}$. This is, ${\bf B}_{\rm{RF}} = B_{\rm{RF}}\ \cos{(\omega_{\rm{RF}}t+\varphi_{\rm{RF}})}\ \hat{x} + B'_{\rm{RF}}\ \cos{(\omega_{\rm{RF}}t+\varphi_{\rm{RF}})}\ \hat{y}$. In this case, under the resonance condition, we  find (in the rotating frame of $H_0 = \frac{\omega}{2} \sigma_z$) 
\begin{equation}
    H'=\frac{\sqrt{\Omega_x^2+\Omega_y^2}}{2} \begin{pmatrix}
        0 & e^{i\left(\varphi_{\rm RF} -\arctan{\frac{\Omega_y}{\Omega_x}}\right)}\\
        e^{-i\left(\varphi_{\rm RF} -\arctan{\frac{\Omega_y}{\Omega_x}}\right)} & 0
    \end{pmatrix},
\end{equation}
which induces rotations along an axis 
\begin{equation}\label{eqA6}
    \hat{k}= \cos{\left(\varphi_{\rm RF} -\arctan{\frac{\Omega_y}{\Omega_x}}\right)} \ \hat{x} + \sin{\left(\varphi_{\rm RF} -\arctan{\frac{\Omega_y}{\Omega_x}}\right)} \ \hat{y}.
\end{equation}
For simplicity, and in concordance with our experimental conditions, we continue with the case $\Omega_y=0$.

The expectation value of the nuclear spin operator ${\bf I}\equiv {\bf \sigma}/2$ is given by  
\begin{equation}
    \langle{\bf I}(t)\rangle = \Tr[\rho_f {\bf I}\ ]= \Tr[U_S^\dagger(t)\ {\bf I}\ U_S(t) \ \rho_0],
\end{equation}
where $\rho_f$ and $\rho_0$ are the final and initial states, respectively and $U_S(t)$ is the evolution operator in the Schrödinger's picture.

For clarity, we first consider the case without RF driving ($\Omega = 0$), for which the spin expectation value reduces to
\begin{equation}
    \langle{\bf I}(t)\rangle = \Tr[U_0^\dagger(t)\ {\bf I}\ U_0(t) \rho_0] = \Tr[e^{iH_0 t}\ {\bf I}\ e^{-iH_0 t}\rho_0].
\end{equation}
Evaluating explicitly the components of ${\bf I}$, we obtain
\begin{eqnarray}
    \langle I_x(t) \rangle &=& \Tr[I_x \ \rho_0] \ \cos{(\omega t)} - \Tr[I_y \ \rho_0] \ \sin{(\omega t)}, \\
    \langle I_y(t) \rangle &=& \Tr[I_x  \  \rho_0] \ \sin{(\omega t)} + \Tr[I_y   \ \rho_0] \ \cos{(\omega t)}, \\
    \langle I_z(t) \rangle &=& \Tr[I_z  \  \rho_0], 
\end{eqnarray}
with $\Tr[I_i \ \rho_0]=\langle I_i\ (t=0)\rangle$ for $i=\{x,\ y, \ z\}$.

Extending this to a spin ensemble, we identify the expectation value $\langle {\bf I}(t)\rangle$ with the  magnetization vector ${\bf M}(t)$, such that
\begin{equation}
    {\bf M}(t) =
\begin{pmatrix}
M_x(t) \\
M_y(t) \\
M_z(t)
\end{pmatrix}
=
\begin{pmatrix}
M_x(0)\cos\omega t - M_y(0)\sin\omega t \\
M_x(0)\sin\omega t + M_y(0)\cos\omega t \\
M_z(0)
\end{pmatrix}.
\end{equation}
We can write the previous result as
\begin{equation}
    {\bf M}(t) = \begin{pmatrix}
\cos\omega t & -\sin\omega t & 0 \\
\sin\omega t & \cos\omega t & 0 \\
0 & 0 & 1
\end{pmatrix}
\begin{pmatrix}
M_x(0) \\
M_y(0) \\
M_z(0)
\end{pmatrix}.
\end{equation}
This explicitly shows that, under a static field, the magnetization precesses around the $z$-axis with angular frequency $\omega$.

When an RF driving pulse is present (i.e., $\Omega \neq 0$), the expectation value of the nuclear spin becomes
\begin{equation}
    \langle {\bf I}  \rangle = \Tr[e^{i\frac{\Omega}{2}\sigma_{\varphi_{\rm{RF}}} t} e^{i\omega\frac{\sigma_z}{2}t}\ {\bf I}\ e^{-i\omega\frac{\sigma_z}{2}t}e^{-i\frac{\Omega}{2}\sigma_{\varphi_{\rm{RF}}} t}\rho_0].
\end{equation}
Performing similar computations, one finds
\begin{equation}\label{appsimplerot}
    {\bf M}(t)=R_{\hat{z}}(\omega t)\ R_{\hat{k}}(\Omega t) \  {\bf M}_0,
\end{equation}
where $R_{\hat{z}}(\omega t)$ and $R_{\hat{k}}(\Omega t)$ are 3-dimensional rotation matrices and represents precession around axes $\hat{z}$ and $\hat{k}= \cos\varphi_{\rm{RF}} \ \hat{x} + \sin\varphi_{\rm{RF}} \hat{y}$. Hence, the RF drive induces an additional rotation about the transverse axis defined by the initial RF phase $\varphi_{\rm{RF}}$. 

The $R_{\hat{z}}(\omega t)$ and $R_{\hat{k}}(\Omega t)$ matrices are 
\begin{equation}
R_{\hat{z}}\left(\omega t\right) = \begin{pmatrix}
\cos\omega t & -\sin \omega t & 0\\
\sin \omega t & \cos \omega t & 0\\
0 & 0 & 1
\end{pmatrix},
\end{equation}
 \begin{equation}
R_{\hat{k}}\left(\Omega t \right) = \\
 \begin{pmatrix}
\cos{\Omega t} - (-1 + \cos{\Omega t}) \cos^2{\varphi_{\rm{RF}}} &  - (-1 + \cos{\Omega t}) \cos{\varphi_{\rm{RF}}} \sin{\varphi_{\rm{RF}}} & \sin{\Omega t} \sin{\varphi_{\rm{RF}}} \\
- (-1 + \cos{\Omega t}) \cos{\varphi_{\rm{RF}}} \sin{\varphi_{\rm{RF}}} & \cos{\Omega t} - (-1 + \cos{\Omega t}) \sin^2{\varphi_{\rm{RF}}} & -\cos{\varphi_{\rm{RF}}} \sin{\Omega t} \\
- \sin{\Omega t} \sin{\varphi_{\rm{RF}}} & \cos{\varphi_{\rm{RF}}} \sin{\Omega t} & 1 - 2 \sin^2{\frac{\Omega t}{2}}
\end{pmatrix},
\end{equation}
where $\varphi_{\rm{RF}}$ corresponds the azimuthal angle in the plane XY (see Fig.~\ref{Fig2} for clarity). 

Employing the rotation formalism for magnetization vectors offers an intuitive yet rigorous framework for a clear description of spin dynamics.

\section{Characterization of magnetic fields sensed by NVs under concurrent MW and RF driving}\label{appB}
\subsection{Sample magnetic field in terms of magnetization vectors}
We consider an NV center located at a depth $d$ of approximately a few nanometers below the diamond surface, beneath a sample containing $N$ spins. Note, the NV detects nuclear spins within a hemispherical volume of radius $d$ above it~\cite{Staudacher2013, Mamin2013}.

The joint NV–nuclei dynamics is governed by 
\begin{equation}
H = D S_z^2 -\gamma_e B_{\rm{ext}} S_z- \sum_{j=1}^{N} \gamma_j B_{\rm{ext}} I_j^z 
+ \sum_{j=1}^{N} \frac{\hbar \mu_0 \gamma_e \gamma_j}{2 |{\bf r}_j|^3} 
\left( {\bf S} \cdot {\bf I}_j - 3 \frac{({\bf S} \cdot {\bf r}_j)({\bf I}_j \cdot {\bf r}_j)}{|{\bf r}_j|^2} \right),
\end{equation}
where an external magnetic field ${\bf B} = B_{\rm{ext}} \; \hat{z}$ is aligned with the NV axis,  $D = (2 \pi)\ \times\ 2.87$ GHz is the zero-field splitting, $\gamma_e = -(2 \pi)\ \times\ 28.8$ GHz/T is the electron gyromagnetic factor, $\gamma_j$ denotes the nuclear gyromagnetic ratio, in our case $\gamma_H=(2 \pi)\ \times\ 42.58$ MHz/T, and $\mu_0$ is the vacuum permeability. In a rotating frame   with respect to $D S_z^2 -\gamma_e B_{\rm{ext}}S_z$, and applying the rotating wave approximation (RWA), the Hamiltonian simplifies to
\begin{equation}
H' = - \sum_{j=1}^{N} \gamma_j B_{\rm{ext}} I_j^z 
+ \sum_{j=1}^{N} S_z \ {\bf A}_j \cdot {\bf I}_j,
\end{equation}
with
\begin{equation}
{\bf A}_j = \frac{\hbar \mu_0 \gamma_e \gamma_j}{2 |{\bf r}_j|^3} 
\left( \hat{z} - 3 \frac{(\hat{z} \cdot {\bf r}_j) \cdot {\bf r}_j}{|{\bf r}_j|^2} \right).
\end{equation}

A second rotating frame transformation  (in this case w.r.t. $- \sum_{j=1}^{N} \gamma_j B_{\rm{ext}} I_j^z $) enables us to write the interaction term among nuclei and the NV center as follows 
\begin{eqnarray}
    \sum_{j=1}^{N} S_z {\bf A}_j \cdot {\bf I}_j 
\;\longrightarrow\;
S_z  \frac{\rho \hbar \mu_0 \gamma_e \gamma_j}{2} 
\langle {\bf I} \rangle 
\cdot \int\left[ g_x(r), g_y(r), f(r) \right] dV,
\end{eqnarray}
where the geometric functions $g_{x,y}(r)$ and $f(r)$ read
\begin{equation}
    g_{x,y}(r) = \frac{1}{r^5}\,(3r_{z} r_{x,y}) \quad {\rm{and}} \quad f(r)=\frac{1}{r^3}\left(\frac{3r_{z}^{2}}{r^2} - 1\right),
\end{equation}
with $\hat{r} = (r_{x}, r_{y}, r_{z})$ being the unitary vector that joins the NV with the effective volume occupied by a single nucleus, and $r$ is their relative distance. In addition, the expectation value $\langle {\bf I} \rangle \equiv 2 {\bf M}(t)$ serves as an effective representation of the nuclei in the semi-bubble on the diamond surface. The parameter $\rho$ represents the nuclear spin density, defined as the number of spin-active nuclei per unit volume, $\rho=N/V$. In practice, it depends on the composition of the sample; for organic samples predominantly composed of hydrogen, $\rho \sim 6 \times 10^{28}$ m$^{-3}$~\cite{Heisse2017}.

In this manner, the interaction term $\sum_{j=1}^{N} S_z \ {\bf A}_j \cdot {\bf I}_j$ can be rewritten in the form of a Zeeman-like coupling, $-\gamma_e S_z b_z(t)$, where $b_z(t)$ denotes the magnetization-dependent magnetic field generated by the sample. This is given by
\begin{equation}
    b_z(t) = \frac{\rho \hbar \mu_0 \gamma_j}{4} {\bf M}(t) 
\cdot \int \left[ g_x(r), g_y(r), f(r) \right] \, dV.
\end{equation}
Due to the geometrical symmetry of the spin ensemble around the NV axis (as illustrated in Fig.~\ref{Fig1}a in the main text), the integral over $g_y(r)$ vanishes. This is because $g_y(r) \propto r_z r_y$ is an odd function with respect to reflections about the $y=0$ plane.  Additionally, although the $z$-component of the integral (arising from $f(r)$) may be nonzero in general, it does not contribute to the signal since the MW driving induces changes in the $S_z$ operator of every NV which are not compensated owing to the static character of that field component. Thus, the magnetic field detected by a single NV simplifies to
\begin{equation}
    b_z(t) = B_{\rm{max}} \ {\bf M}(t) \cdot \hat{x}.
\end{equation}
For clarity, we can group the constants into a single prefactor $B_{\rm{max}}$ whose value can be determined from the experimental setup and includes the geometric factor computed from the integral.

\subsection{Magnetic fields detected by the NV during combined MW and RF sequence}
The system Hamiltonian for a single NV reads 
\begin{equation}
H = DS_z^2 - \gamma_e B_{\rm{ext}} S_z - \gamma_e\,b_z(t) S_z + \sqrt{2} \,\Omega\, S_x \cos\left(\omega_{\rm{MW}} t -\varphi_{\rm{MW}}\right),
\end{equation}
where $\Omega_{\rm{MW}}$ and $\omega_{\rm{MW}}$ are the amplitude and frequency of the MW control driving.

When an RF pulse is applied to the sample, the magnetic field detected by the NV center is modified as a consequence of the change in the magnetization of the sample, a first example of the latter is first described in Eq.~(\ref{appsimplerot}). Note that, in our protocol, the RF pulse is applied before the correlation time stage, as illustrated in Fig.~\ref{Fig1}c. 

During the 1st interrogation block, the magnetic field experienced by the NV is given by
\begin{equation}\label{eq_BS_int}
    b_1(t) = B_{\rm{max}} \ {\bf M}(t) \cdot \hat{x} = B_{\rm{max}} \, R_{\hat{z}}(\omega t) \; {\bf M}_0\cdot \hat{x} = B_{\rm{max}}\; \sin{\alpha} \cos{(\omega t + \beta)},
\end{equation}
for $0\leq t<t_1$. Here, $\omega$  is the Larmor frequency of the sample nuclei, ${\bf M}_0$ is the initial magnetization vector, defined as
\begin{equation}
    {\bf M}_0=\begin{pmatrix}\sin{\alpha} \cos{\beta}\\
    \sin{\alpha} \sin{\beta}\\
    \cos{\alpha}
    \end{pmatrix},
\end{equation}
with $\alpha\in [0,\pi]$ and $\beta \in [0, 2\pi]$ are the polar and azimuthal, as depicted in Fig.~\ref{Fig1}b.

While the RF pulse is applied, the magnetization vector precesses around both the $\hat{z}$-axis and the RF driving ``axis" $\hat{k}$, which is $\hat{k} = \cos{\varphi_{\rm{RF}}}\hat{x} + \sin{\varphi_{\rm{RF}}}\hat{y}$. As a result of this double precession in this interaction picture, the magnetic field detected by the NV during the RF pulse is given by
\begin{eqnarray}
    b_2(t) &=& B_{\rm{max}}\,R_{\hat{z}}(\omega t) \,R_{\hat{k}}(\Omega t) \,R_{\hat{z}}(2 \omega \tau) \, {\bf M}_0\cdot \ \hat{x} \\
    &=& B_{\rm{max}}\,R_{\hat{z}}(\omega t) \,R_{\hat{k}}(\Omega t)\,{\bf M}_1\cdot \ \hat{x},
\end{eqnarray}
for $t_1 \leq t < t_2$, where ${\bf M}_1=R_{\hat{z}}(2 \omega \tau) \, {\bf M}_0$ is the magnetization vector immediately after the 1st interrogation block of duration $2\tau$ (see Fig.~\ref{Fig1}c), and $\omega$ is the nuclear Larmor frequency. 

During the  correlation time and the 2nd interrogation block stage, the magnetic field is
\begin{eqnarray}
    b_3(t) &=& B_{\rm{max}}\ R_{\hat{z}}(\omega t)\ R_{\hat{z}}(\omega t_p) \ R_{\hat{k}}(\Omega t_p) \,R_{\hat{z}}(2\omega \tau) \ {\bf M}_0\cdot \ \hat{x}, \\
    &=& B_{\rm{max}}\ R_{\hat{z}}(\omega t)\ R_{\hat{z}}(\omega t_p) \ R_{\hat{k}}(\Omega t_p) \  {\bf M}_1\cdot \ \hat{x}, \\
    &=& B_{\rm{max}}\ R_{\hat{z}}(\omega t)\  {\bf M}_2\cdot\ \hat{x}\ ,
\end{eqnarray}
for $t \geq t_2$. Here $t_p = t_2-t_1$ is the duration of the RF drive and ${\bf M} _2=R_{\hat{z}}(\omega t_p) \, R_{\hat{k}}(\Omega t_p) \, {\bf M}_1$ is the magnetization vector right after the RF pulse stage.

\section{Accumulated phases during the  protocol}\label{appC}
The sequence in Fig.~\ref{Fig1}c, including both the MW and the RF channel, leads to the propagator
\begin{equation}
U = e^{-i\frac{\pi}{4}\sigma_x} \, \sigma_y\, e^{-i\frac{\phi_4}{2}\sigma_z} \,e^{-i\frac{\pi}{4}\sigma_y} \,
e^{-i\frac{\phi_2}{2}\sigma_z} \,e^{-i\frac{\phi_3}{2}\sigma_z} \,
e^{-i\frac{\pi}{4}\sigma_x} \, \sigma_y \, e^{-i\frac{\phi_1}{2}\sigma_z} \,e^{-i\frac{\pi}{4}\sigma_y}.
\end{equation}
Where the phases $\phi_1$, $\phi_2$, $\phi_3$, and $\phi_4$ correspond to those accumulated during the first interrogation stage, the RF driving stage, the free evolution period, and the second interrogation stage, respectively. They can be computed as follows
\begin{eqnarray}
\phi_1 & = & \int_0^{\tau} b_1(s) \; {\rm{ds}} - \int_\tau^{2\tau} b_1(s) \; {\rm{ds}},\label{eq_phi1}\\
\phi_2 &=& \int_{2\tau}^{2\tau +t_p} b_2(s) \; {\rm{ds}}, \label{phi2}\\
\phi_3 &=& \int_{2\tau +t_p}^{2\tau +t_p +\tilde{\tau}} b_3(s) \; {\rm{ds}},\label{phi3}\\
\phi_4 &=& \int_{2\tau + t_p +\tilde{\tau}}^{3\tau + t_p +\tilde{\tau}} b_3(s) \; {\rm{ds}} - \int_{3\tau + t_p + \tilde{\tau}}^{4\tau + t_p +\tilde{\tau}} b_3(s) \; {\rm{ds}},\label{phi4}
\end{eqnarray}

The corresponding magnetic fields, before assuming the resonance condition, are given by
\begin{eqnarray}
    b_1(t) &=& B_{\rm{max}} \cos(\beta + \omega t)\, \sin(\alpha),\\
    b_2(t) &=& B_{\rm{max}} \Big[ 
\cos(\alpha)\, \sin(\varphi_{\rm{RF}} + \omega t)\, \sin(\Omega t)+ \frac{1}{2}\sin(\alpha)\,\Big(
\cos[\beta - 2\varphi_{\rm{RF}} - \omega t + 2\tau\omega]\nonumber\\
&&
+ \cos[\beta + \omega t + 2\tau\omega]- \sin[\beta - \varphi_{\rm{RF}} + 2\tau\omega]\,
\big(\sin[\varphi_{\rm{RF}} + \omega t - \Omega t]
+ \sin[\varphi_{\rm{RF}} + \omega t + \Omega t]\big)
\Big)\Big],\\
b_3(t) &=& B_{\rm{max}} \Big[
\cos\!\big(\beta + (\omega t + \omega t_p + 2\omega\tau)\big)
\cos^2\!\left(\frac{\Omega t_p}{2}\right) \sin(\alpha)\nonumber\\
&&+ \cos\!\big(\beta - 2\varphi_{\rm{RF}} - \omega t - \omega t_p + 2\omega\tau\big)
\sin^2\!\left(\frac{\Omega t_p}{2}\right) \sin(\alpha) + \cos(\alpha)\, \sin[\varphi_{\rm{RF}} + \omega(t + t_p)]\, 
\sin(\Omega t_p)
\Big].
\end{eqnarray}

Thus, the accumulated phases are as follows 
\begin{eqnarray}
\phi_1 & = & \frac{4\,B_{\rm{max}}|\gamma_e|}{\omega} \sin^2 \left(\frac{\omega \tau}{2}\right)\sin (\alpha) \sin \left(\omega \tau + \beta\right),\\
\phi_2&=& \tfrac{1}{2} B_{\rm max}\Bigg[\frac{2\cos\!   \left(\beta - 2\varphi_{\rm RF} - \tfrac{t_p \omega}{2}\right)\sin\alpha\;\sin\!\left(\tfrac{t_p \omega}{2}\right)}{\omega}+ \frac{2\cos\!\left(\beta + \tfrac{t_p \omega}{2} + 4\tau\omega\right)\sin\alpha\;\sin\!\left(\tfrac{t_p \omega}{2}\right)}{\omega}\nonumber\\
&&+ \frac{\cos\alpha\,|\gamma_e|}{(\omega-\Omega)(\omega+\Omega)}\Big((\omega+\Omega)\,\sin\!\big[\varphi_{\rm RF} + (t_p + 2\tau)(\omega-\Omega)\big]- (\omega+\Omega)\,\sin\!\big[\varphi_{\rm RF} + 2\tau\omega - 2\tau\Omega\big]\nonumber\\
&&+ (\omega-\Omega)\,\big(\sin\!\big[\varphi_{\rm RF} + 2\tau(\omega+\Omega)\big]- \sin\!\big[\varphi_{\rm RF} + (t_p + 2\tau)(\omega+\Omega)\big]\big)\Big)\nonumber\\
&&+ \frac{ \sin\alpha\,\sin(\beta-\varphi_{\rm RF}+2\tau\omega)}{\omega-\Omega}\,\Big(\cos\!\big[\varphi_{\rm RF} + (t_p+2\tau)(\omega-\Omega)\big]- \cos\!\big[\varphi_{\rm RF} + 2\tau\omega - 2\tau\Omega\big]\Big) \nonumber \\
&&+ \frac{ \sin\alpha\,\sin(\beta-\varphi_{\rm RF}+2\tau\omega)}{\omega+\Omega}\,\Big(- \cos\!\big[\varphi_{\rm RF} + 2\tau(\omega+\Omega)\big]+ \cos\!\big[\varphi_{\rm RF} + (t_p+2\tau)(\omega+\Omega)\big]\Big)\Bigg],\\
\phi_3 &=&\frac{B_{\rm max}\,|\gamma_e|}{\omega}\Bigg[
\cos^2\!\left(\frac{t_p \Omega}{2}\right)\sin\alpha\,
\Big(-\sin(\beta + 2 t_p \omega + 4\tau\omega)
+ \sin\big(\beta + (2 t_p + 4\tau + \tilde{\tau})\omega\big)\Big)
\nonumber\\
&&+ 2\,\cos\!\Big(\beta - 2\varphi_{\rm RF} - \tfrac{(4 t_p + \tilde{\tau})\omega}{2}\Big)\,
\sin\alpha\,\sin\!\Big(\tfrac{\tilde{\tau}\,\omega}{2}\Big)\,
\sin^2\!\left(\frac{t_p \Omega}{2}\right)\nonumber\\
&&+ \cos\alpha\,
\Big(
\cos\!\big[\varphi_{\rm RF} + 2(t_p+\tau)\omega\big]
- \cos\!\big[\varphi_{\rm RF} + (2(t_p+\tau)+\tilde{\tau})\omega\big]
\Big)\,
\sin(t_p \Omega)
\Bigg],\\
\phi_4 &=& \frac{B_{\rm max}\,|\gamma_e|}{\omega}\Bigg[
- \cos^2\!\left(\frac{t_p \Omega}{2}\right)\sin\alpha\,
\Big(
\sin\big[\beta + (2 t_p + 4\tau + \tilde{\tau})\omega\big]
- 2 \sin\big[\beta + (2 t_p + 5\tau + \tilde{\tau})\omega\big]+ \sin\big[\beta + (2 t_p + 6\tau + \tilde{\tau})\omega\big]
\Big)\nonumber\\
&&+ \sin\alpha\,\sin^2\!\left(\frac{t_p \Omega}{2}\right)
\Big(
\sin\big[\beta - 2\varphi_{\rm RF} - (2 t_p + \tilde{\tau})\omega\big]
- 2 \sin\big[\beta - 2\varphi_{\rm RF} - (2 t_p + \tau + \tilde{\tau})\omega\big]+ \sin\big[\beta - 2\varphi_{\rm RF} - (2(t_p+\tau) + \tilde{\tau})\omega\big]
\Big)\nonumber\\
&&+ \cos\alpha\,\sin(t_p \Omega)\,
\Big(
-2 \cos\big[\varphi_{\rm RF} + (2 t_p + 3\tau + \tilde{\tau})\omega\big]
+ \cos\big[\varphi_{\rm RF} + (2 t_p + 4\tau + \tilde{\tau})\omega\big]+ \cos\big[\varphi_{\rm RF} + (2(t_p+\tau) + \tilde{\tau})\omega\big]
\Big)
\Bigg].
\end{eqnarray}

The previous equations can be simplified when assuming resonance condition ($2\tau=1/\omega)$ leading to some of the expressions in the main text.
\begin{eqnarray}
b_1(t) &=& B_{\rm{max}}\; \sin{\alpha} \cos{(\omega t + \beta)},\\
b_2(t) &=& B_{\text{max}} \left[ \sin{( \Omega t)} \ \cos{(\alpha)}\ \sin{(\varphi_{\rm{RF}} + \omega t)}\right. \nonumber\\
&&\left. +\ \frac{1}{2}\sin{(\alpha)}\ \left(\cos{(\beta-2\varphi_{\rm{RF}}-\omega t)} + \cos{(\omega t + \beta)} - \sin{(\beta-\varphi_{\rm{RF}})}\left(\sin{(\varphi_{\rm{RF}}+\omega t - \Omega t)} + \sin{(\varphi_{\rm{RF}}+\omega t + \Omega t)}\right)\right) \right], \nonumber \\ \\
b_3(t) &=& B_{\text{max}}  \left[ \cos{(\Omega\ t_p/2)}^2\ \sin{(\alpha)} \ \cos{(\omega\ (t+t_p)+\beta)} \right.\nonumber \\
&&\left.+ \sin{(\Omega\ t_p/2)}^2\ \sin{(\alpha)} \ \cos{(\beta-2\ \varphi_{\rm{RF}}-\omega(t+t_p))}+ \sin{(\Omega\ t_p)}\ \cos{(\alpha)}\ \sin{(\varphi_{\rm{RF}}+\omega(t+t_p))} \right].
\end{eqnarray}

\begin{eqnarray}
    \phi_1 &=& -\,\frac{4 B_{\rm{max}}\,|\gamma_e|}{\omega}\,\sin\alpha\,\sin\beta ,\\
\phi_2 &=& \tfrac{1}{2} B_{\rm{max}} \Bigg[
\frac{
\big(
\cos[\varphi_{\rm{RF}} + (t_p + \tfrac{2\pi}{\omega})(\omega - \Omega)]
- \cos[\varphi_{\rm{RF}} - \tfrac{2\pi\Omega}{\omega}]
\big)
\sin\alpha\,\sin(\beta - \varphi_{\rm{RF}})
}{\omega - \Omega}\nonumber\\
&&+ 
\frac{
\big(
-\cos[\varphi_{\rm{RF}} + \tfrac{2\pi\Omega}{\omega}]
+ \cos[\varphi_{\rm{RF}} + (t_p + \tfrac{2\pi}{\omega})(\omega + \Omega)]
\big)
\sin\alpha\,\sin(\beta - \varphi_{\rm{RF}})
}{\omega + \Omega}+
\frac{2\cos[\beta - 2\varphi_{\rm{RF}} - \tfrac{t_p\omega}{2}]\,\sin\alpha\,\sin(\tfrac{t_p\omega}{2})}{\omega}
\nonumber\\
&&
+ \frac{2\cos[\beta + \tfrac{t_p\omega}{2}]\,\sin\alpha\,\sin(\tfrac{t_p\omega}{2})}{\omega}
+
\frac{|\gamma_e|\,\cos\alpha}{(\omega - \Omega)(\omega + \Omega)} \Big[
(\omega + \Omega)\,\sin[\varphi_{\rm{RF}} + (t_p + \tfrac{2\pi}{\omega})(\omega - \Omega)]
- (\omega + \Omega)\,\sin[\varphi_{\rm{RF}} - \tfrac{2\pi\Omega}{\omega}]
\nonumber\\
&&+ (\omega - \Omega)
\big(
\sin[\varphi_{\rm{RF}} + \tfrac{2\pi\Omega}{\omega}]
- \sin[\varphi_{\rm{RF}} + (t_p + \tfrac{2\pi}{\omega})(\omega + \Omega)]
\big)
\Big]
\Bigg],
\end{eqnarray}

\begin{eqnarray}
    \phi_3 &=& \frac{2 B_{\rm{max}}\,|\gamma_e|}{\omega}\,
\sin\!\left(\frac{\tilde{\tau}\omega}{2}\right)
\Big[
\cos\!\left(\beta + 2t_p\omega + \frac{\tilde{\tau}\omega}{2}\right)
\cos^2\!\left(\frac{t_p\Omega}{2}\right)\sin\alpha
+
\cos\!\left(\beta - 2\varphi_{\rm{RF}} - \tfrac{(4t_p + \tilde{\tau})\omega}{2}\right)
\sin\alpha\,\sin^2\!\left(\frac{t_p\Omega}{2}\right)
\nonumber\\
&&+
\cos\alpha\,
\sin\!\left(\varphi_{\rm{RF}} + 2t_p\omega + \tfrac{\tilde{\tau}\omega}{2}\right)
\sin(t_p\Omega)
\Big] ,\\
\phi_4 &=& \frac{4 B_{\rm{max}}\,|\gamma_e|}{\omega}\,
\Big[
- \cos^2\!\left(\frac{t_p\Omega}{2}\right)\sin\alpha\,
\sin(\beta + 2t_p\omega + \tilde{\tau}\omega)
+
\sin\alpha\,\sin^2\!\left(\frac{t_p\Omega}{2}\right)\,
\sin(\beta - 2\varphi_{\rm{RF}} - (2t_p + \tilde{\tau})\omega)
\nonumber\\
&&+
\cos\alpha\,
\cos(\varphi_{\rm{RF}} + 2t_p\omega + \tilde{\tau}\omega)\,
\sin(t_p\Omega)
\Big].
\end{eqnarray}

\section{Experimental Methods}
\subsection{Sample Preparation}
A $2 \times 2 \times 0.5$\,mm electronic-grade diamond crystal ($^{12}$C-enriched, Fraunhofer IAF, Freiburg) was implanted with nitrogen at an energy of 2.5\,keV, using a 7$^\circ$ off-axis tilt and a fluence of $2 \times 10^{12}$\,cm$^{-2}$ by Innovion, and subsequently annealed following the protocol described by Bucher \textit{et al.}\,\cite{Bucher2019}. Before the experiments, the diamond was cleaned using the triacid cleaning protocol described by Brown \textit{et al.}\,\cite{Brown2019}. A drop of silicon oil (Sigma-Aldrich, CAS 63148-62-9) was then placed on the diamond surface to provide a proton-rich sample for demonstrations of nuclear spin manipulation.

\subsection{Correlation Experiment}
Correlation spectroscopy measurements were performed using two XY8-4 dynamical decoupling blocks separated by a variable correlation interval $(\tilde \tau)$. Each experimental cycle included a 5\,\(\mu\)s laser pulse for spin initialization and readout, and MW \(\pi/2\) and \(\pi\) pulses of 20\,ns and 40\,ns duration, respectively. The \(\ket{0}\leftrightarrow\ket{-1}\) transition was addressed at 1.996 \, GHz, corresponding to a static magnetic field of approximately \(B_0 \approx 31.2\) mT. In this field, the proton Larmor frequency is 1.33 \, MHz. The interpulse delay was therefore set to \(\tau = 188.1\) ns, matching the detection frequency to the proton Larmor frequency.
For nuclear spin manipulation, an RF pulse was inserted between the two XY8-4 blocks and the correlation time $\tilde \tau$ was swept in 51 points from 60\,\(\mu\)s to 63\,\(\mu\)s. For each RF amplitude, a full correlation trace was acquired over this interval.  The start phase of the RF pulse was adjusted to compensate for the delay between the beginning of the pulse sequence and the onset of the RF pulse, such that the desired effective initial phase was established at the moment of nuclear spin driving. Each trace was recorded with 6000 averages and repeated over 3 sweeps.

\subsection{Experimental Setup}

The experiments were conducted on the $^{12}$C-enriched diamond crystal glued onto a thin glass coverslip. A \SI{6}{\milli\meter} diameter glass hemisphere (TECHSPEC\textsuperscript{\textregistered} N-BK7 Half-Ball Lens, Edmund Optics) was affixed to the underside of the coverslip to enhance fluorescence collection efficiency. The coverslip was mounted on a \SI{30}{\milli\meter} cage plate (CP4S, Thorlabs), and the entire assembly was positioned between two permanent magnets generating a static magnetic field of approximately \SI{31}{\milli\tesla}, aligned along one of the four NV axes.

Initialization and readout of the NV spin state were performed using a laser (Verdi G7, Coherent) operating at a power of approximately \SI{200}{\milli\watt}. Laser pulses of \SI{5}{\micro\second} duration were generated using an acousto-optic modulator (3250-220, Gooch and Housego). After passing through a $\lambda/2$ waveplate to optimize the polarization for NV excitation, the laser beam was focused onto the diamond using a lens in a total internal reflection geometry. The resulting NV fluorescence was collected and collimated by a condenser lens (ACL25416U-B, Thorlabs), filtered with a long-pass filter (Edge Basic 647 Long Wave Pass, Semrock), and directed through a second identical condenser lens onto an avalanche photodiode (ACube S3000-10, Laser Components). The photodiode output was recorded with a data acquisition unit (USB-6229 DAQ, National Instruments).

MW pulses were synthesized directly by a 25 GS/s (AWG70000, Tektronix) and amplified using a \SI{16}{\watt} power amplifier (ZHL-16W-43-S+, Mini-Circuits). Direct synthesis of the MW control avoided intermediate mixing stages or an external microwave source, thereby preserving phase coherence with the AWG-defined timing reference. MW signals were delivered to the diamond through a home-built loop antenna fabricated from a shortened coaxial cable\cite{Bucher2019}.

RF pulses for nuclear spin manipulation were generated by a separate RF source (DG4162, Rigol), amplified (LZY-22+, Mini-Circuits), and delivered through a second loop positioned concentrically above the MW loop. Both loops were aligned parallel to the diamond surface to ensure efficient coupling of the MW and RF fields to the NV and nuclear spins, respectively.

The timing of all components was controlled by the AWG, which acted as the master clock of the experiment. The acousto-optic modulator was triggered by the AWG to gate the laser pulses, while the DAQ system was synchronized to detect the NV fluorescence in the appropriate readout windows. For correlation experiments with nuclear spin control, the RF generator was run in externally triggered gated burst mode. An AWG trigger set the timing and duration of the RF pulse so that it was applied only in the control window between the two XY8 blocks. The RF phase was adjusted to set the desired phase relation to the MW reference at pulse onset.
\end{document}